\documentclass[epsfig,12pt]{article}
\usepackage{epsfig}
\usepackage{graphicx}
\usepackage{array}
\usepackage{color}
\usepackage{bbm}
\usepackage{bm}
\graphicspath{{img/}}
\usepackage{subcaption,float}
\usepackage{amsmath}
\usepackage{amssymb} 
\usepackage{csquotes} 
\usepackage{physics}
\usepackage{slashed}
\usepackage{mathtools}
\usepackage{amsfonts}
\usepackage{threeparttable}
\usepackage{adjustbox}
\usepackage[nottoc,notlof,notlot]{tocbibind} 
\usepackage[titles]{tocloft} 
\usepackage{booktabs}
\usepackage[justification=centering]{caption}
\usepackage[T1]{fontenc}
\usepackage{xcolor}
\usepackage{hyperref}
\usepackage[numbers,sort&compress]{natbib}

\usepackage{cleveref}

\newcommand{\beq}{\begin{equation}}   
\newcommand{\eeq}{\end{equation}}
\newcommand{\beqn}{\begin{eqnarray}}   
\newcommand{\eeqn}{\end{eqnarray}}

\newcommand{\pt}{\partial}

\newcommand{\gsim}{\lower.7ex\hbox{$
\;\stackrel{\textstyle>}{\sim}\;$}}
\newcommand{\lsim}{\lower.7ex\hbox{$
\;\stackrel{\textstyle<}{\sim}\;$}}
\usepackage[textsize=scriptsize]{todonotes}
\def\bar{\overline}
\def\tilde{\widetilde}
\def\d{\partial}
\def\vp{\varphi}
\def\RR{\mathbb{R}}
\def\ZZ{\mathbb{Z}}
\def\cH{\mathcal{H}}
\def\cE{\mathcal{E}}

\begin{document}

\begin{titlepage}

\begin{flushright}
FTPI-MINN-26-10,  UMN-TH-2528/26\\
UCI-TR-2026-04\\

\end{flushright} 

\vspace{2mm}

\begin{center}
	{\large{\bf
	More on Classical Stability of Hopf-like Solitons of the\\[1mm] Toroidal-Twisted type	}}

\vspace{3mm}
	
{  \bf Chao-Hsiang Sheu$^a$ and Mikhail Shifman$^b$}

\end{center}
\begin{center}
{\it $^a$Department of Physics and Astronomy, University of California,
Irvine, CA 92697-4575}\\[5pt]
{\it  $^b$ William I. Fine Theoretical Physics Institute,
University of Minnesota,
Minneapolis, MN 55455}\\
\end{center}

\vspace{5mm}

\begin{center}
{\large\bf Abstract}
\end{center}

The Faddeev-Hopf model \cite{FN}  supporting  Hopfions was shown to emerge in the low-energy limit of four-dimensional scalar 
quantum electrodynamics (QED) with two charged scalar fields \cite{2,3}. Faddeev and Noemi conjectured that the Hopfions and Hopf-like solitons -- vortons -- can be based on a twisted toroidal structure inherent to QED (\cite{Davis:1988jp,Davis:1988jq,Radu:2008pp}). This conjecture was discussed in detail in \cite{3} in the approximation of negligibly small extrinsic curvature. Qualitative and semi-quantitative arguments were used to demonstrate the validity of the Faddeev-Noemi hypothesis. Here we further enhance the proof by applying a numerical analysis which confirms that  { large-size} Hopf-like solitons exist as local energy
minima in the full QED theory  (in the Faddeev-Skyrme model they become
topological solitons representing the global minima in the given topological sector). 
\end{titlepage}

\section{Introduction}

The model with the Hopf solitons originally introduced  by Faddeev and Niemi \cite{FN} is a deformed O(3) nonlinear sigma
model in {\em four dimensions}, which in our notation takes the form
\begin{equation}
S =\int d^4 x\left[ \frac{\xi}{4}\,\, \partial_{\mu}\vec{S}
\,\partial^{\mu}\vec{S} -\frac{\beta -1 }{4g^2}\,F_{\mu\nu} F^{\mu\nu} 
\right]\,,
\label{senza}
\end{equation}
where the three-component field $\vec S$ is an ``isotopic''  vector subject to the constraint
\begin{equation}
\vec S^{\,2} =1\,,  
\label{ts}
\eeq
and 
\beq
F_{\mu\nu} =\frac 1 2  \varepsilon_{abc} S^a\pt_\mu S^b\pt_\nu S^c
\eeq
The second term in (\ref{senza}) presents a deformation of the O(3) model. 
Sometimes it is referred to as the Faddeev-Hopf model,
for a review see \cite{manton}.
The constant $\xi$ has dimension $[m^2]$ while $\beta$ and $g^2$ are dimensionless. 
Since we will be interested only in static solitons, the time coordinate is irrelevant, only three spatial dimensions
$x,y$ and $z$ enter our consideration.

The 
vacuum of the model (\ref{senza}) corresponds to a constant value of $|\vec S|$ which can be
chosen arbitrarily, say,  as $(\vec S )_{\rm vac} =(0,0,1).$ Due to (\ref{ts}) the target space of the
sigma model at hand is $S_2$.
Finiteness of the soliton energy  requires that the vector $\vec S$ 
must tend to its vacuum value at the spatial infinity, namely,
\begin{equation}
\vec S\to \{0,0,1\}\,\,\,\mbox{at}\,\,\, \left| \vec x \right| \to\infty\,.
\label{vac}
\end{equation}
The boundary condition (\ref{vac})
compactifies the space to $S_3$. Since $$\pi_3 (S_2) = \ZZ \,,$$
the knot solitons present topologically nontrivial maps of $S_3\to S_2$.
As was noted in \cite{FN}, there is an associated integer topological
number, the Hopf invariant.

The Hopf invariant cannot be written as an integral of a local density of the 
field $\vec S$. However, if one uses a U(1)
gauged formulation of the CP(1) $\sim$ O(3) sigma model 
(for a review see \cite{nsvzrev,uch} and a detailed list of references in \cite{FN}), then the Hopf invariant ${\cal H}$
reduces to the Abelian Chern-Simons term for a gauge field in the above CP(1) gauged representation,
\beq
\mathcal{H} = \frac{1}{4\pi ^{2}}\int dxdydz\,\epsilon ^{\mu \nu \rho }
\left( A_{\mu}\partial _{\nu }A_{\rho }\right).
\label{tue-two}
\eeq

\begin{figure}[h]
\epsfxsize=10cm
\centerline{\epsfbox{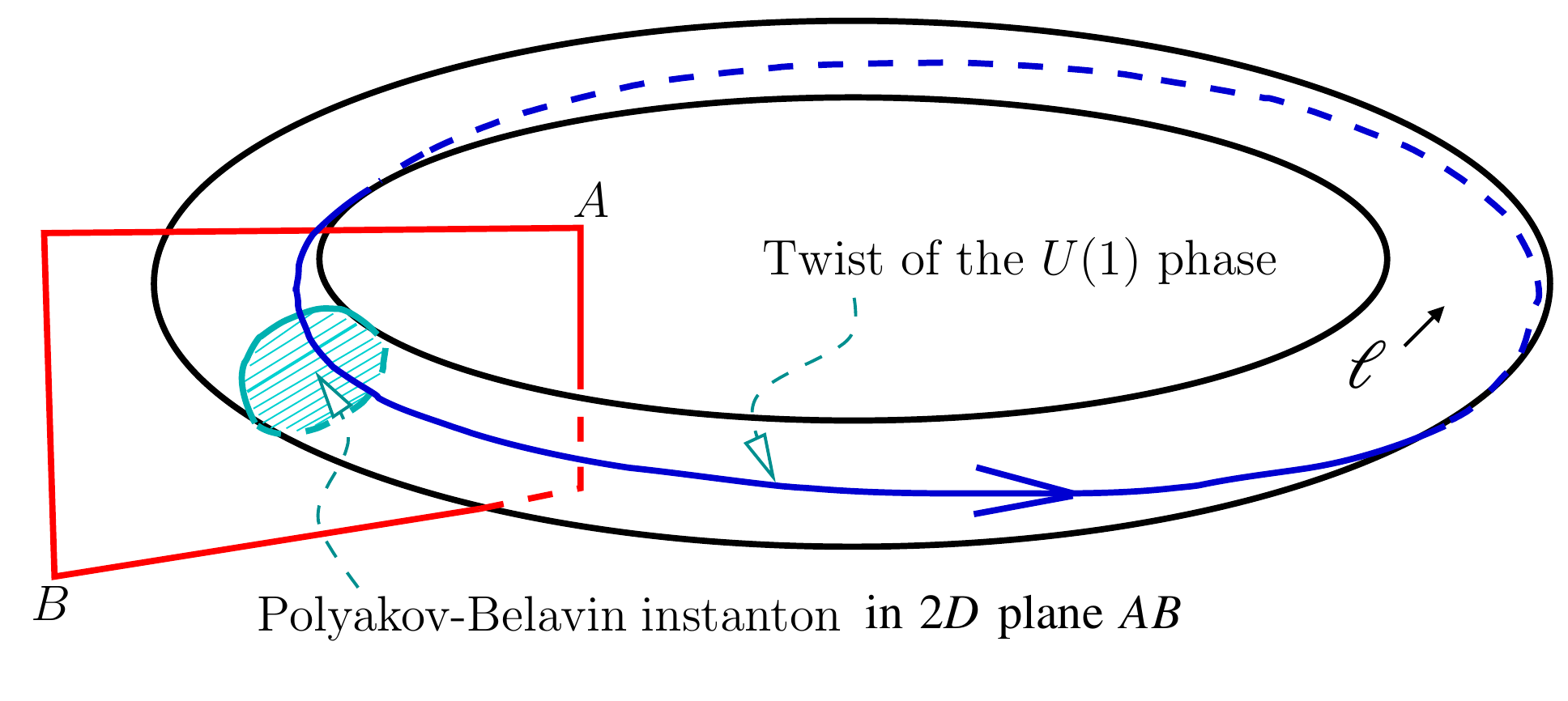}}
\caption{\footnotesize The simplest Hopf-like soliton as a string with the $\alpha$ twist
illustrating the Faddeev-Niemi hypothesis \cite{FN}.  A Belavin-Polyakov ``instanton'' on the plane $AB$ is extended in  one extra dimension
$\ell$ and bent into a torus, with   a $2\pi$ twist of the instanton  phase modulus.}
\label{fig1}
\end{figure}

\begin{figure}[h]
\epsfxsize=10cm
\centerline{\epsfbox{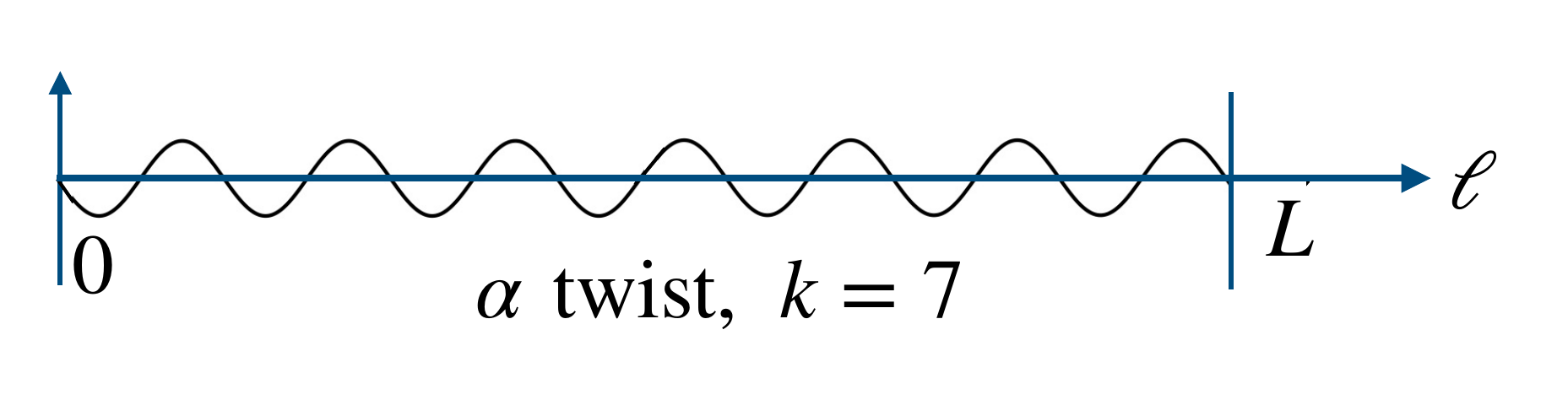}}
\caption{\footnotesize The twist of the angular modulus, with $k=7$.}
\label{fig2}
\end{figure}

According to  \cite{FN}, the simplest Hopf-like soliton, which we study here in the so-called vorton approximation, can be visualized  as a vortex tube bent in the form of a twisted torus, see Fig. \ref{fig1}.
A sketch in this figure can be interpreted as a two-dimensional topological soliton in the  $AB$ plane, with a topological charge $\sum_{\alpha\beta=1,2} \int dx dy  \varepsilon ^{\alpha\beta}F_{\alpha\beta}(x,y) $, supplemented by the twist number $\int dz  A^z$, combine to produce a nonvanishing Hopf-like topological invariant (\ref{tue-two}). In the model (\ref{senza}) the soliton in the $AB$ plane is the Polyakov-Belavin instanton which has a complex modulus $\rho = |\rho |e^{i\alpha}$ where $|\rho |$ is the size modulus while $\alpha$ is the an{g}ular modulus which can wind along the large period of the torus $L_\ell$ (Fig. \ref{fig2}),
\beq
A_z= \frac{2\pi k}{L_\ell}\,,\quad \alpha = \frac{2\pi k z}{L_\ell}\,.
\eeq 
We assume that the circumference of the small period of the torus $L_s\ll L_\ell$. Then the extrinsic curvature of the toroidal soliton which tends to shrink it  is small. 

Following \cite{3}, we will simplify the problem. Instead of the torus in $\RR^3$ with size $L_\ell$ along the larger cycle (see Fig. \ref{fig1}) we will consider $\RR^2\times S^1$ with $r(S^1)= L_{s}$. Then the torus with an extrinsic curvature\footnote{The extrinsic curvature of torus becomes negligible in the limit $L_\ell\gg L_s$.} is converted into a cylinder with the vanishing extrinsic curvature and a nonvanishing Hopf-like invariant. 

We pause here to clarify the difference between the {\em bona fide} Hopf invariant and its close ``relative,'' the Hopf-like invariant we will be dealing with.

Assume the axis of the cylinder mentioned above is aligned with the $z$ axis. The perpendicular $x,y$ plane hosts the cross section of the cylinder tube (in this case
the Polyakov-Belavin ``instanton''). 
Consider a Hopf-like invariant defined as 
\begin{align}
\label{eq:hopflike}
  h \stackrel{\rm def}{=} \frac{1}{2\pi^2}\int_{0}^{L}\dd{z}\int\dd{x}\dd{y} \epsilon^{ijl}F_{ij}\d_{l}\alpha\,.
\end{align}
Given the twisted string ansatz \eqref{eq:Sinphi} and \eqref{eq:Ainphi} (see below), $h$ equals  $kn$, i.e. the product of the twist $k$ and the 2D topological number $n$. The former is the winding number of the $U(1)$ phase twist along the longitudinal $z$-direction, while the latter is the vortex winding number in the perpendicular $xy$-plane.

The relationship between  $h=kn$ defined in \cref{eq:hopflike} and  the Hopf invariant \eqref{tue-two} is obvious, and so is the difference between them.
Indeed, recall that the Hopf invariant of a map $S^3\to S^2$ counts the linking number of pre-image curves of two distinct points on $S^2$~\cite{FN,manton,Shnir:2018yzp}. 
In the CP(1) parametrization \eqref{eq:nparam} (see Sec. \ref{2two}), the field $\vec{S}$ is constructed from two complex homogeneous coordinates $n_A$ whose phases wind along two orthogonal cycles, namely, the azimuthal angle $\theta$ in the transverse plane and the longitudinal coordinate $z$ along the string axis (the $S^{1}$ direction).
Then \cref{eq:hopflike} can be realized as the product of integrands over two orthogonal cycles.
The transverse magnetic flux $\int \dd{x}\dd{y}\epsilon^{ij}F_{ij}$ measures the vortex winding in the $xy$-plane while the longitudinal phase gradient $\int_0^L \dd{z}\d_z\alpha$ counts the twist along the string axis.
This matches with the notion of counting the linking number.

The distinction from the {\em bona fide} Hopf invariant is topological. 
Our construction is defined on $\mathbb{R}^2\times S^1$ rather than on (a compact) $S^3$, so the homotopy 
classification $\pi_3(S^2)=\mathbb{Z}$ is  not directly applicable. The  {\em bona fide}  Hopf invariant vanishes on this geometry.
This can be seen by a direct computation of $\cH$ with the ansatz \eqref{eq:Ainphi}.

The soliton  under consideration belongs to the class of  the so-called
{\em vortons},  solutions first constructed in \cite{Davis:1988jp,Davis:1988jq} (see also~\cite{Radu:2008pp} for a review).
In the vorton picture, a vortex string carrying an internal current, the twist of the phase modulus $\alpha(z)$, is bent into a loop.
The charge $h=kn$ plays the role of the vorton charge $N=nm$ defined in \cite{Radu:2008pp}.
It is classically conserved and prevents the unwinding of the twist.
In fact, we regard $h=kn$ as a topological number of the vorton. The winding $n$ is protected by the magnetic flux from $\pi_1(S^1)=\ZZ$, while the longitudinal winding $k$ characterizes the twist along the $z$-cycle. Hence $h=kn$ is stable against small fluctuations. This is the meaning of the  vorton genuine topological quantum numbers, even though $\pi_3(S^2)$ is not directly applicable on $\RR^2\times S^1$.

As was described in \cite{3} and references therein, the study of Hopfions in various 4D models,  which reduce to (\ref{senza}) in the low-energy limit, started long ago, both in condensed matter and high-energy physics. One of the most popular models is the 4D QED with two or more charged matter fields.
The most serious problem in this approach is to demonstrate dynamical  $L_{s,\ell}$ stabilization of the soliton solution. The issue was addressed in Ref. \cite{3} where the analytic stability analysis
was based on a number of approximations (see references in \cite{3} for previous work). In this paper we complement analytic arguments on the dynamical stability by numerical calculations in the vorton approximation.

Before wrapping up this introduction, we make one comment that connects our straight twisted-string construction to the toroidal vorton. If we bend the twisted string into a torus ring of circumference $L_\ell$ such that
\begin{align}
L_\ell \;\gg\; L_s \;\sim\; |\rho|\,,
\end{align}
the resulting toroidal configuration in $\RR^3$ has extrinsic curvature suppressed by  powers of $\sim 1/L_\ell$. This need not necessarily  destabilize it: 
the twisted string will survive as a local minimum configuration.  The extrinsic-curvature-induced bending force will eventually destroy the vorton ring through tunnellng, but the associated decay rate will  be exponentially small. Consequently, the configuration remains \emph{quasi-stable}.
It is well-localized in field space, carries a conserved Hopf-like charge $h=kn$, and decays only on time scales much longer than the natural dynamical scales of the theory. In this sense the toroidal Hopf-like soliton may be viewed as the large-loop limit of a twisted tube  with no extrinsic curvature.

The paper is organized as follows. In Sec. \ref{sec:model} we present the model suitable for the analysis of twisted semilocal vortex tubes.
In this section we introduce the twist $k$ and topological number $n$ of the 2D vortex.
Section \ref{sec:numrics} is devoted to our numerical investigation on the stability of the twisted vortex tube. We discuss in detail the numerical scheme and demonstrate that the string is dynamically stabilized at an equilibrium length $L_*$.
Section \ref{sec:compare} compares our results with related numerical and analytic work in the literature.

\section{Model for a twisted semilocal vortex \label{sec:model}}
\label{2two}

Before getting into the numerical results, we present the setup for the twisted semilocal string solution and the associated physical quantities. We restrict attention to static configurations.

From the action \eqref{senza}, the energy functional of a time-independent configuration takes the form 
\begin{align}\label{eq:energy}
\cE = \int\dd[3]{x} \left(  
  \frac{\xi}{4}\d_{i}\vec{S}\cdot\d_{i}\vec{S} + \frac{\beta-1}{4g^2}F_{ij}F_{ij}
\right)
\end{align} 
where $i,j=1,2,3$.
To construct the twisted semilocal-string ansatz, recall that we can parametrize the isotopic vector $\vec{S}$ and the gauge field $A_{\mu}$ using the homogeneous coordinates $n_{A}$ of the CP(1) model. Namely, 
\begin{align}\label{eq:nparam}
  S^{a} = \bar{n}_{A}\tau^{a}n^{A}
  \,,\qquad
  A_{\mu} = -\frac{i}{2}\left( \bar{n}_{A}\stackrel{\leftrightarrow}\partial_{\mu}n^{A} \right)
\end{align}
where $a=1,2,3$, $A=1,2$ and $\tau^{a}$ are the Pauli matrices.
With the U(1) gauge freedom and the constraint on $n_{A}$, the semilocal-string ansatz can be expressed as
\begin{subequations}\label{eq:Sinphi}
\begin{align}
  S_1 =&~ 2\xi^{-1} \varphi(r)\sqrt{\xi - \varphi^2(r)}\cos[n\theta+\alpha(z)]
  \\[2mm]
  S_2 =&~ 2\xi^{-1} \varphi(r)\sqrt{\xi - \varphi^2(r)}\sin[n\theta+\alpha(z)]
  \\[2mm]
  S_3 =&~ 2\xi^{-1}\varphi^2(r) - 1
\end{align}\end{subequations}
for $\vec{S}$ and 
\begin{align}
\label{eq:Ainphi}
    A_1(\vec{x}) =  -n\xi^{-1}\varphi^2(r) \frac{y}{r^2} 
    \,,\quad
    A_2(\vec{x}) =  n\xi^{-1}\varphi^2(r) \frac{x}{r^2},\nonumber\\[2mm]
    A_3(\vec{x}) =  \xi^{-1}\varphi^2(r) \alpha'(z)
\end{align}
for the gauge field, in terms of a scalar field $\varphi(r)$ (the real part of $n_{1}$) and a compact twisting parameter $\alpha(z)$ along the $z$ direction. Here $n$ labels the winding around the azimuthal angle $\theta$.
We work in cylindrical coordinates $(r, \theta, z)$ with the string axis aligned along $z$.
In addition, the boundary condition for $\vp(r)$ is 
\begin{align}\label{eq:phibc}
    \varphi(0) = 0 \,,\quad \varphi(\infty) = \sqrt{\xi} 
    \,.
\end{align}

Substituting \cref{eq:Sinphi,eq:Ainphi} in \cref{eq:energy} gives the energy functional, 
\begin{multline}\label{eq:energyr}
\cE = \int_{0}^{L}\dd{z}\int_{\RR^2}\dd{r}\dd{\theta} \Biggl\{
  \frac{\vp^2}{\xi r} \left( \xi - \varphi^2 \right)\left( n^2+r^2\alpha'^2 \right)
  + \frac{\xi r \varphi'^2}{\xi - \varphi^2}
  \\
  + \frac{2(\beta-1)}{g^2\xi^2 r}\varphi^2\varphi'^2\left( n^2+r^2\alpha'^2 \right)
\Biggr\}
\end{multline}
The equations of motion can be derived accordingly for $\alpha$
\begin{align}\label{eq:eomalpha}
  \pdv[2]{\alpha(z)}{z} = 0 
\end{align}
and for $\vp$,
\begin{align}\label{eq:eomvarphi}
0 =&
\left[ \frac{4(\beta-1)}{g^2\xi^2r}\varphi^2\left( n^2+r^2\alpha'^2 \right) 
    + \frac{2\xi r}{\xi-\vp^2}
    \right]\varphi''
    \nonumber\\[2mm] 
    &+ \frac{2\varphi}{\xi^2 r}\left[  
        \frac{2(\beta-1)}{g^2}\left( n^2+r^2\alpha'^2 \right)
        + \frac{\xi^3r^2}{(\xi-\varphi^2)^2}
    \right]\varphi'^2
    \nonumber\\[2mm]
    &+\left[  
        \frac{4(\beta-1)}{g^2\xi^2r^2}\varphi^2\left( -n^2+r^2\alpha'^2 \right)
        + \frac{2\xi }{\xi-\vp^2}
    \right]\varphi'
    - \frac{2\varphi}{\xi r} \left( \xi-2\varphi^2 \right)\left( n^2+r^2\alpha'^2 \right).
\end{align}
From \cref{eq:eomalpha} we deduce that $\alpha$ depends on $z$ linearly. Here we consider a $k$-twisted solution,
\begin{align}\label{eq:alphasoln}
  \alpha(z) = \frac{2\pi kz}{L} 
\end{align}
for $k\in\ZZ$ such that the $\vec{S}$ vector is periodic along the $z$ direction with the period $L$. Note that $k$ is the winding number in $z$ and the parameter $L$ can be viewed as the length of this twisted string solution.

At the same time, Eq. (\ref{eq:eomvarphi}) is highly non-linear.  It has to be solved numerically, which will discuss in the next section.

\section{Numerical analysis \label{sec:numrics}}

In this section, we present numerical results on the classical stability of the vorton solution. Varying the length of the twisted string yields a minimum, which indicates the stabilization of the classical vortex solution.

\subsection{Numerical approach}
For simplicity, we focus on the case of $n=1$. Without loss of 
generality, we set $\xi=1$, $g^2=1$, and $k=10$.
Combining  Eqs. \eqref{eq:energyr} and \eqref{eq:alphasoln}, one 
can show that changing $k$ does not alter the solution $\vp(r)$ 
provided the length $L$ is simultaneously rescaled.  However, this rescaling  does 
change the total energy. Namely, under the transformation 
$k \to ak$ and  $L \to aL$ the total energy changes as follows:   $\cE \to a\cE$; 
the functional form of $\vp(r)$ stays intact.

Since we are interested in classical stability, the above 
rescaling does not affect the qualitative picture.  We fix the twist number at  $k=10$ in what follows.

Our numerical procedure to solve \cref{eq:eomvarphi} with the boundary condition \ref{eq:phibc} is described as follows. We compactify the spatial domain via the change of variables
\begin{align}
  r = \frac{x}{1-x}
\end{align}
for $x\in[0,1)$. 
Our numerical solver discretizes the differential equation \eqref{eq:eomvarphi} using a central fourth-order 5-point finite difference stencil to approximate the spatial derivatives. The resulting system of nonlinear algebraic equations is solved using the \verb|root| function from the \verb|scipy.optimize| module.

For a given value of $\beta$ (see eq. (\ref{eq:energy})), we scan the energy across a range of twisted semilocal string lengths using an iterative seeding method. 
We begin by selecting an initial length $L_0$ for which the numerical solution converges efficiently. This solution is then used as a seed to solve for nearby lengths $L_0 \pm \delta L$. Each converged solution serves as the seed for the subsequent step, progressing iteratively to $L_0 \pm 2\delta L$, $L_0 \pm 3\delta L$, and so forth.
Through this procedure, we trace the evolution of the profile functions and the corresponding energy in the vicinity of the minimum.

To ensure numerical convergence, we systematically vary the number of grid points on the interval $[0,1)$ from $300$ to $1000$, verifying that the solution remains stable across different discretizations. The accuracy of the numerical solution is validated by computing the residuals obtained upon substituting the solution back into the discretized equations. These residuals are consistently found to be of order $\mathcal{O}(10^{-5})$, confirming the reliability of our results. The entire computational framework is implemented in 
Python.

\subsection{Results}

\begin{figure}[t]
    \centering
    \begin{subfigure}[b]{0.47\textwidth}
    \centering
    \includegraphics[width=\linewidth]{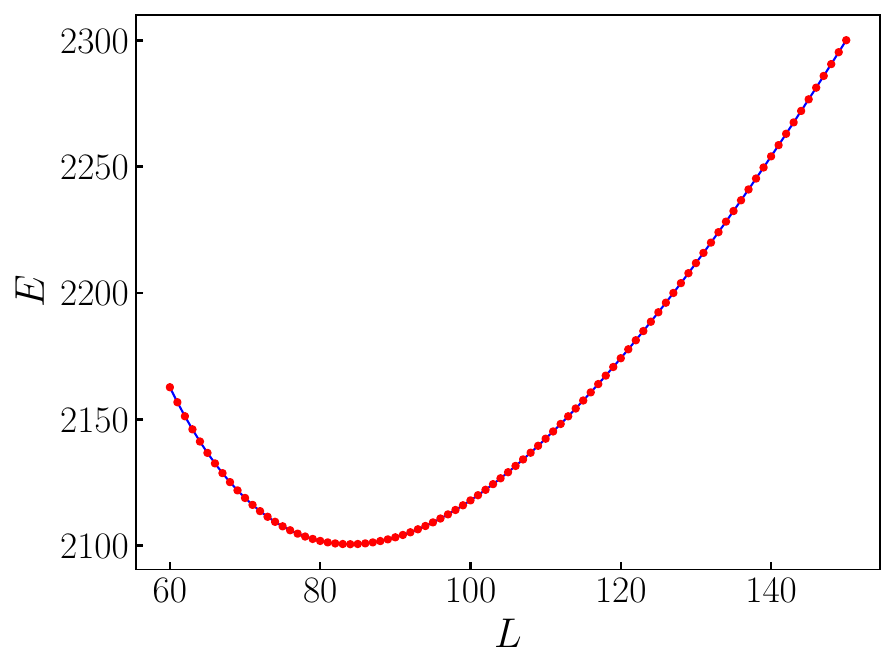}
    \vskip-2ex
    \caption{\footnotesize $\beta=10$}
    \label{fig:EL_beta10m10}
    \end{subfigure}
    \hfill
    \begin{subfigure}[b]{0.47\textwidth}
    \centering
    \includegraphics[width=\linewidth]{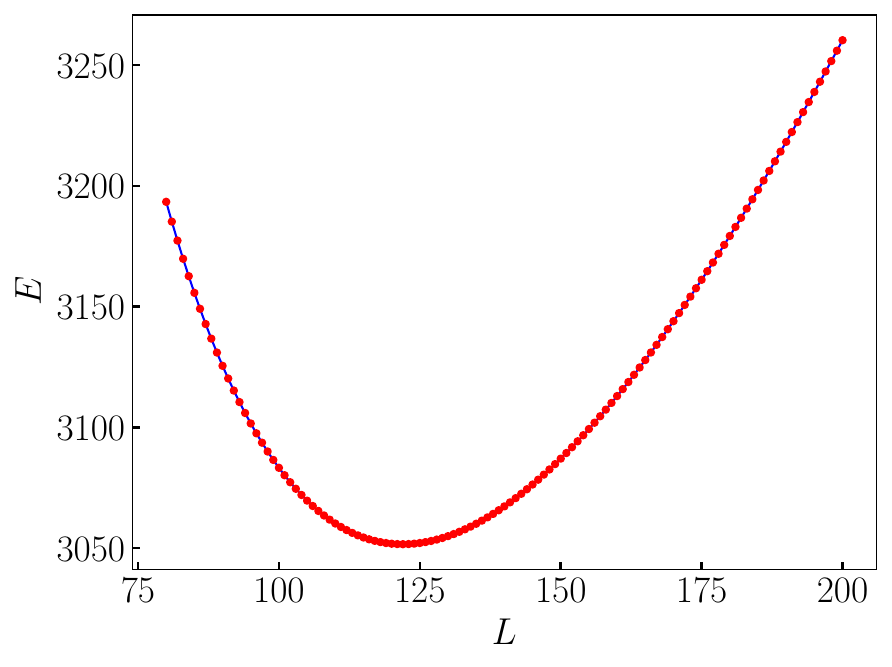}
    \vskip-2ex
    \caption{\footnotesize $\beta=20$}
    \label{fig:EL_beta20m10}
    \end{subfigure}
    \\[3mm]
    \begin{subfigure}[b]{0.47\textwidth}
    \centering
    \includegraphics[width=\linewidth]{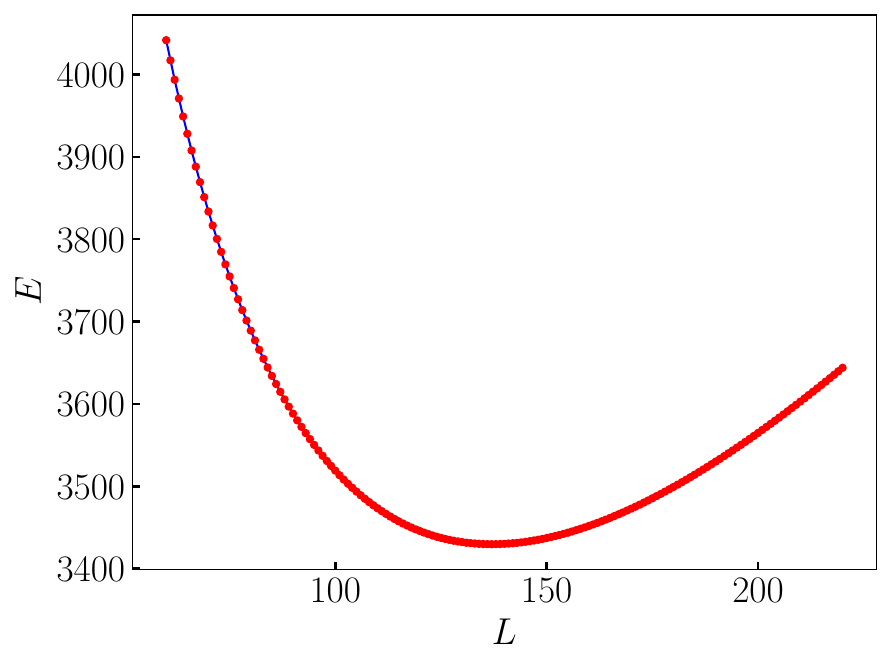}
    \vskip-2ex
    \caption{\footnotesize $\beta=25$}
    \label{fig:EL_beta25m10}
    \end{subfigure}
    \hfill
    \begin{subfigure}[b]{0.47\textwidth}
    \centering
    \includegraphics[width=\linewidth]{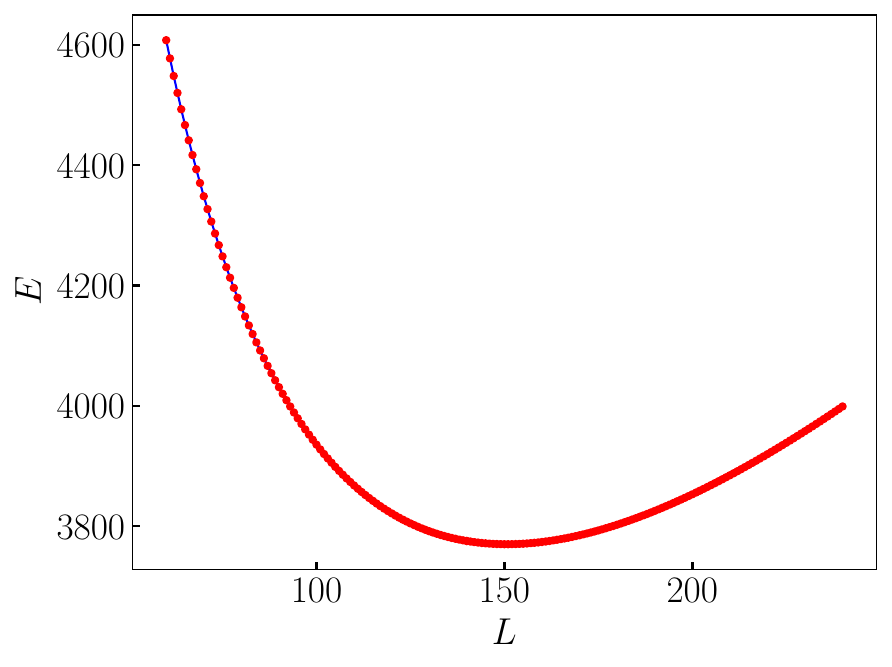}
    \vskip-2ex
    \caption{\footnotesize $\beta=30$}
    \label{fig:EL_beta30m10}
    \end{subfigure}
    \vskip-1ex
    \caption{\footnotesize String energy versus the length of the string for $n=1, g=1, m=10$.}
    \label{fig:ELplots}
\end{figure}

The main result of our numerical analysis is presented in \cref{fig:ELplots}, which displays the total energy $E$ of the twisted semilocal string as a function of its length $L$ for several values of the parameter $\beta$. In all four cases, $\beta = 10, 20, 25$, and $30$, the energy functional exhibits a minimum at a finite value $L = L_*(\beta)$. 
This minimum represents a classically stable configuration where perturbations that attempt to stretch or compress the string along the $z$-direction encounter a restoring force that drives the system back toward $L_*$. 
This is in agreement with the qualitative picture described in \cite{3,Ireson:2018bdw}, namely,
\begin{align}
  \cE(L) \sim \frac{a_{-1}}{L} + a_0 + a_{1}L + \cdots \,.
\end{align}
Comparing  four subplots in \cref{fig:ELplots}, we observe that the total string energy increases systematically with increasing $\beta$, reflecting the enhanced contribution of the latter term in the energy functional \eqref{eq:energy}. 
The existence of such a minimum is the indicator of dynamical stabilization and constitutes the numerical evidence supporting the existence of the stable vorton solution.

From \cref{fig:ELplots}, we observe that the equilibrium length $L_*$ increases monotonically with $\beta$. For $\beta = 10$, the minimum occurs near $L_* \approx 90$, while for $\beta = 30$, it shifts to $L_* \approx 150$. This trend can be understood qualitatively from the energy functional \eqref{eq:energyr} --  larger values of $\beta$ enhance the contribution of the last term proportional to $(\beta - 1)$, which is positive definite, hence pushing the equilibrium to longer strings.
Besides, since the twisted semilocal vortex is stabilized at $L_*$, we identify the soliton mass as $M_{\rm sol} \equiv \cE(L_*)$. The resulting dependence on $\beta$ is displayed in \cref{fig:sltnmass}. The mass grows monotonically with $\beta$, which is consistent with the fact that at larger $\beta$, the dominant $\beta$-dependent contribution in \eqref{eq:energyr} is proportional to $(\beta-1)$.

\begin{figure}[t] 
  \centering 
  \includegraphics[width=.7\linewidth]{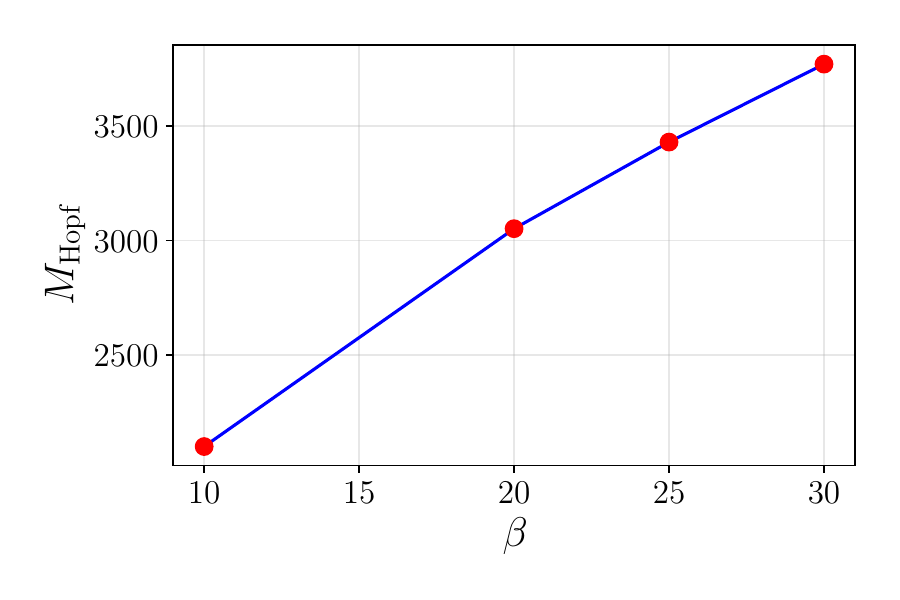}
  \vskip-4ex
    \caption{\footnotesize Soliton masses for $\beta=10,20,25$, and $30$.}
  \label{fig:sltnmass}
\end{figure}

The asymptotic behavior of the $\cE(L)$ curves can also be extracted from the terms in the curly brackets in \cref{eq:energyr}. 
In the limit $L \to 0$, the twist rate $\alpha' \to \infty$, causing the twist-dependent terms in \eqref{eq:energyr} to diverge. 
Note that we expect a vortex profile function should fall in the range $0 < \varphi(r) < \sqrt{\xi}$ in the internal domain $(0,\infty)$. Also, such solution has to be smooth and therefore $\varphi'$ is finite.
This accounts for the steep rise of the energy at small $L$. 
Conversely, as $L \to \infty$, the twist contribution diminishes, but the total energy grows linearly with $L$.  Due to the extensive nature of the string tension  the twist-induced energy that favors large $L$ and the string tension that favors small $L$, produce an energy minimum at intermediate $L_*$.

Moreover,
we have numerically verified  that the Hopf-like invariant \eqref{eq:hopflike} is saturated on our numerical solutions. With the parameters $n = 1$ and $k = 10$ used throughout our analysis, the expected value is $h = kn = 10$. By evaluating the integral given in \cref{eq:hopflike} using the numerical profile functions, we obtain $h = 10$ to an accuracy of order $10^{-4}$ for all values of $\beta$ and $L$ examined. This confirms that the ans\"atze in \cref{eq:Sinphi,eq:Ainphi} preserve the Hopf-like structure and, equivalently, that our numerical solutions carry the expected vorton topological charge $h=kn$ to high accuracy.

\begin{figure}[t]
    \centering
    \begin{subfigure}[b]{0.47\textwidth}
    \centering
    \includegraphics[width=\linewidth]{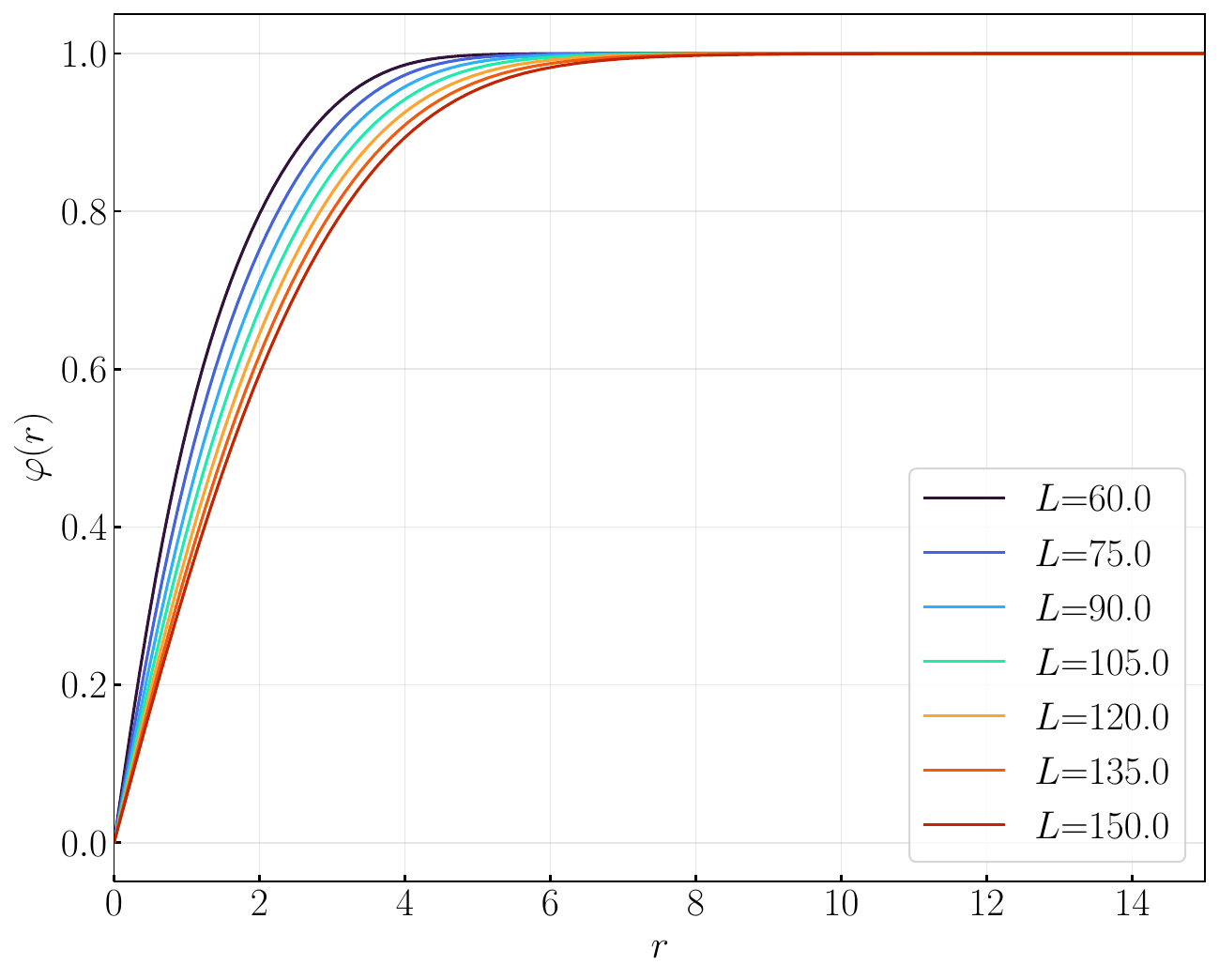}
    \vskip-2ex
    \caption{\footnotesize $\beta=10$}
    \label{fig:wv_beta10m10}
    \end{subfigure}
    \hfill
    \begin{subfigure}[b]{0.47\textwidth}
    \centering
    \includegraphics[width=\linewidth]{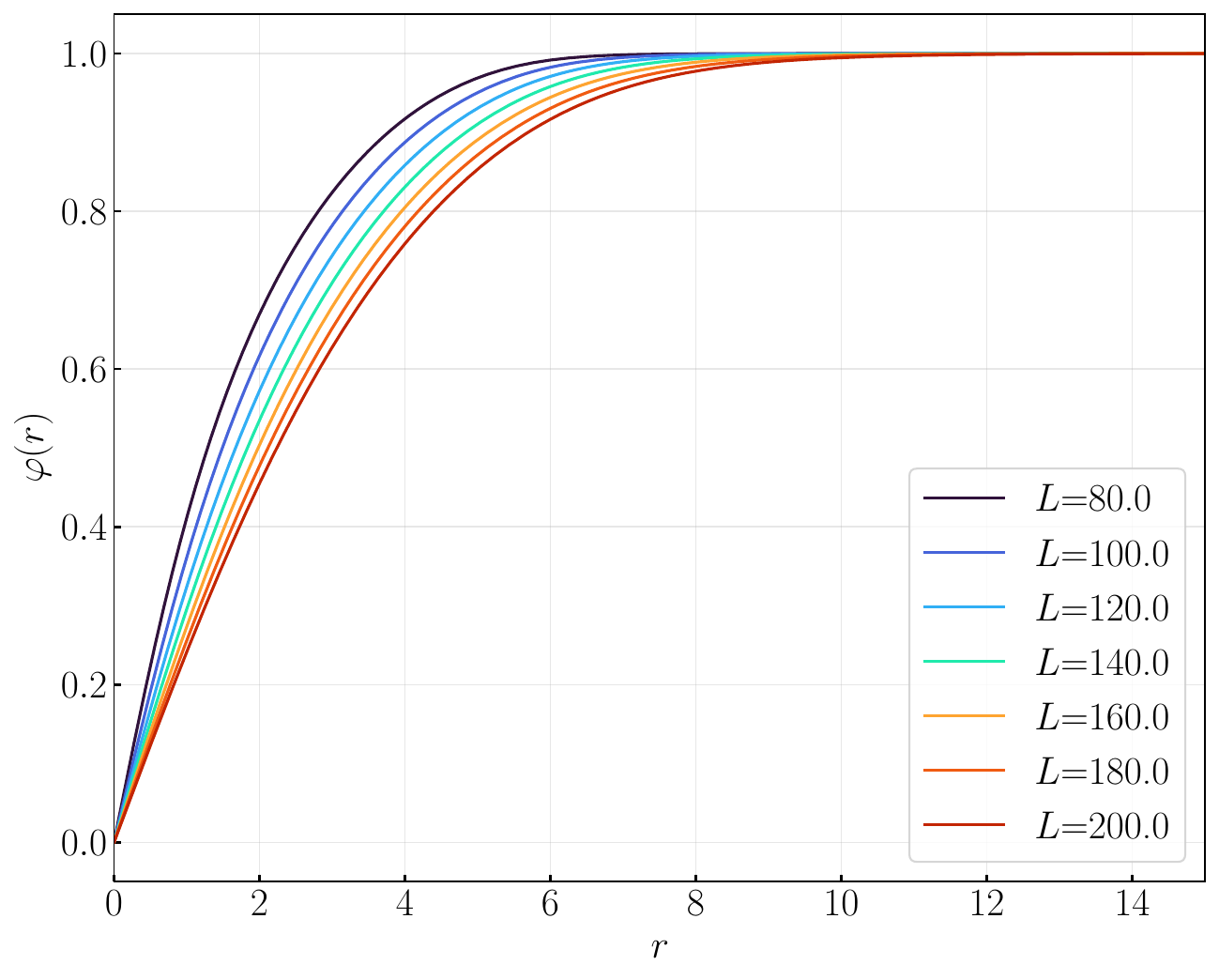}
    \vskip-2ex
    \caption{\footnotesize $\beta=20$}
    \label{fig:wv_beta20m10}
    \end{subfigure}
    \\[3mm]
    \begin{subfigure}[b]{0.47\textwidth}
    \centering
    \includegraphics[width=\linewidth]{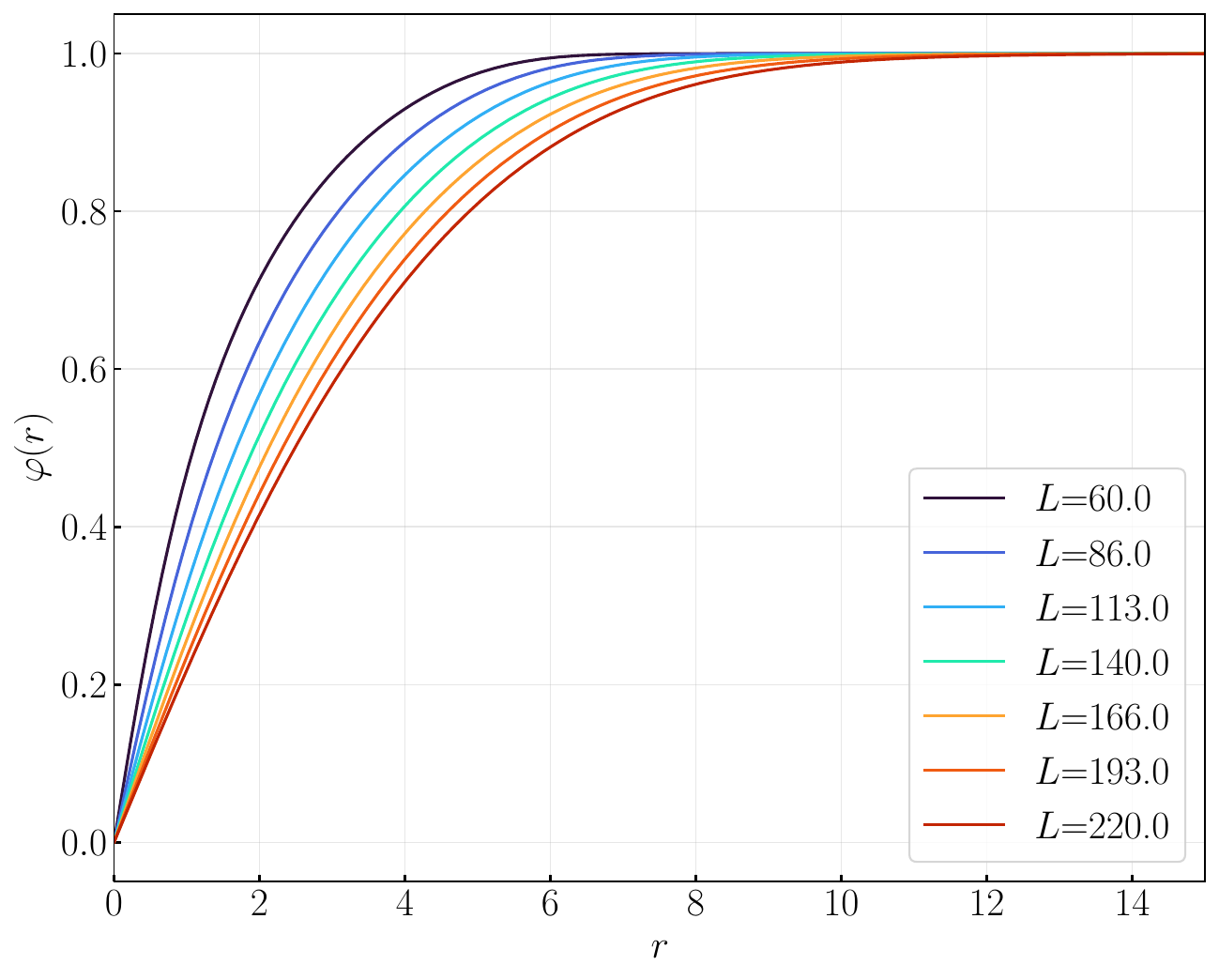}
    \vskip-2ex
    \caption{\footnotesize $\beta=25$}
    \label{fig:wv_beta25m10}
    \end{subfigure}
    \hfill
    \begin{subfigure}[b]{0.47\textwidth}
    \centering
    \includegraphics[width=\linewidth]{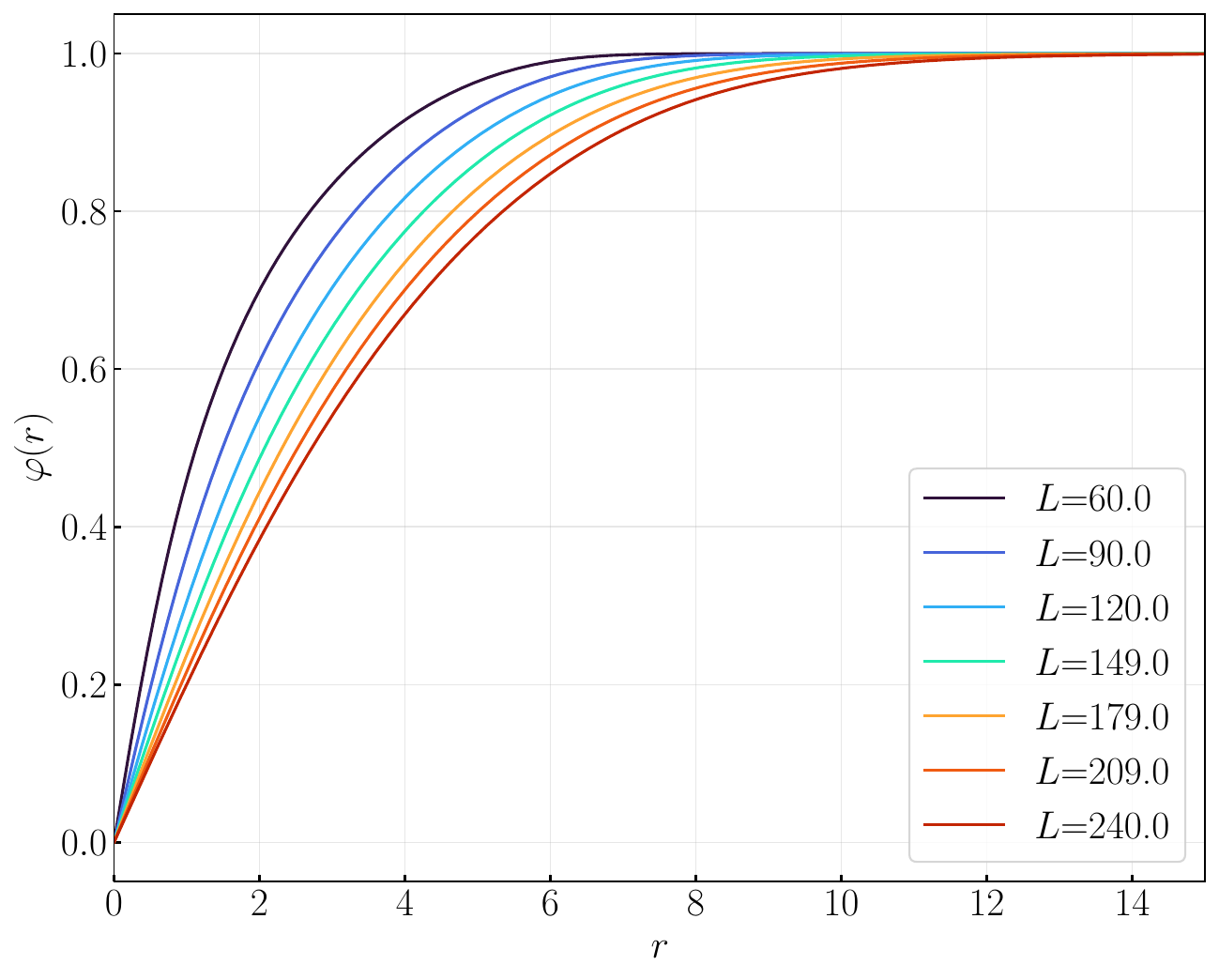}
    \vskip-2ex
    \caption{\footnotesize $\beta=30$}
    \label{fig:wv_beta30m10}
    \end{subfigure}
    \vskip-1ex
    \caption{\footnotesize Profiles of the twisted vortex for $n=1, g=1, m=10$}
    \label{fig:wvplots}
\end{figure}

The profile functions $\vp(r)$ obtained from our numerical solutions are demonstrated in \cref{fig:wvplots} for various string lengths $L$. In all cases, the scalar field satisfies the boundary conditions \eqref{eq:phibc}, vanishing at the origin and approaching $\sqrt{\xi} = 1$ at spatial infinity. The profiles exhibit a characteristic vortex structure, with $\vp(r)$ rising steeply from zero in the core region ($r \lesssim 10$) and approaching  the vacuum value indicated by the minimum of the potential term, i.e. the first term in \cref{eq:energyr}.
This also shows that the string width  $\abs{\rho} \sim 5$ is consistent with the scale hierarchy observed in Ref. \cite{3}. Namely,
\begin{align}
  L_* \gg \abs{\rho} \gg \frac{1}{m_{\gamma}} = \frac{1}{\sqrt{2\xi}g} = \frac{1}{\sqrt{2}}\,.
\end{align}
This is a very important circumstance supporting analytic estimates carried out in \cite{3}.

Notably, for a given $\beta$, profiles corresponding to different $L$ values remain qualitatively similar, differing primarily in the core region $r \sim 1$ to $10$. Shorter strings tend to have slightly broader cores, as the increased twist rate effectively stiffens the radial profile through the coupling in \eqref{eq:eomvarphi}.

\begin{figure}[t]
    \centering
    \begin{subfigure}[b]{0.47\textwidth}
    \centering
    \includegraphics[width=\linewidth]{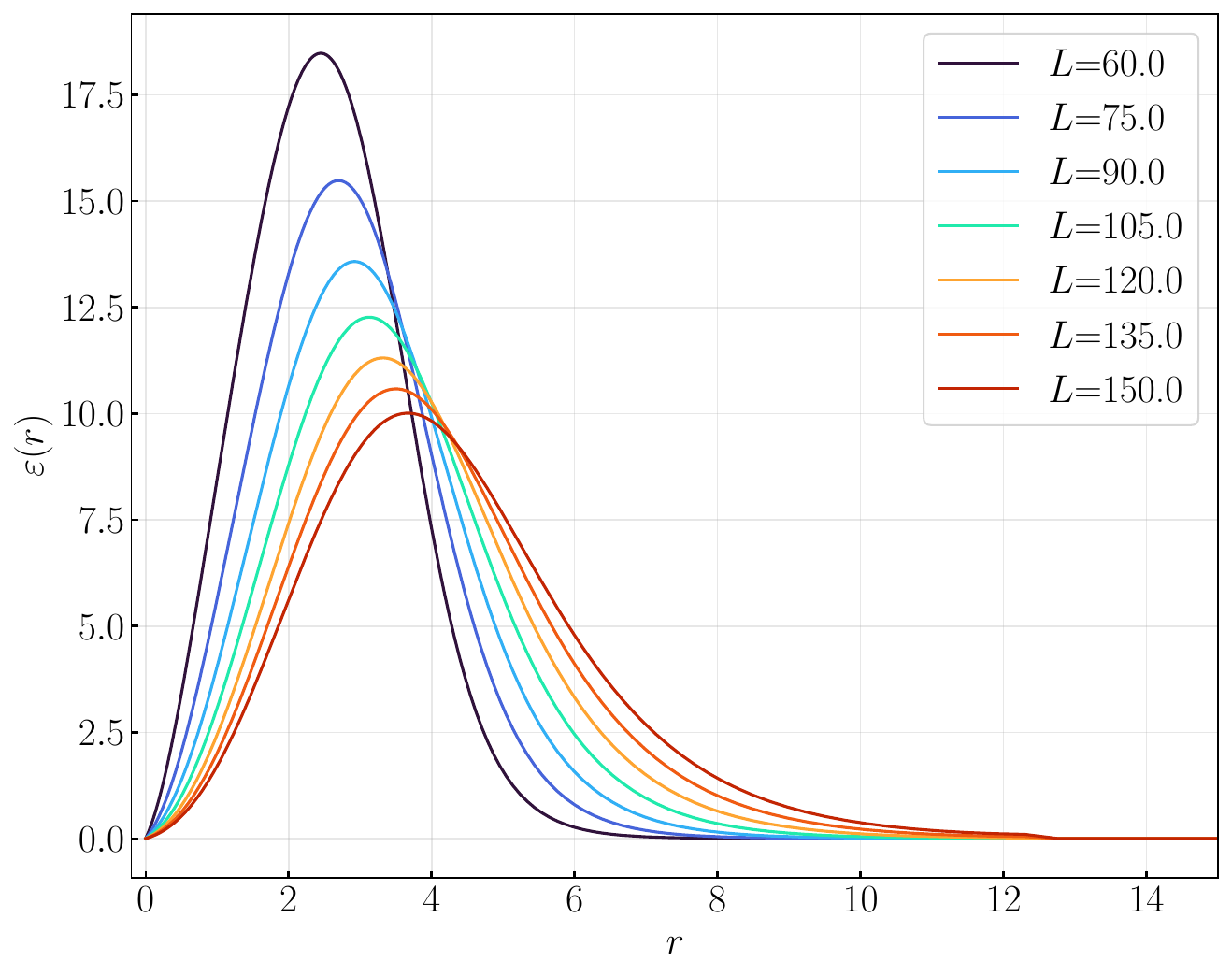}
    \vskip-2ex
    \caption{\footnotesize $\beta=10$}
    \label{fig:ED_beta10m10}
    \end{subfigure}
    \hfill
    \begin{subfigure}[b]{0.47\textwidth}
    \centering
    \includegraphics[width=\linewidth]{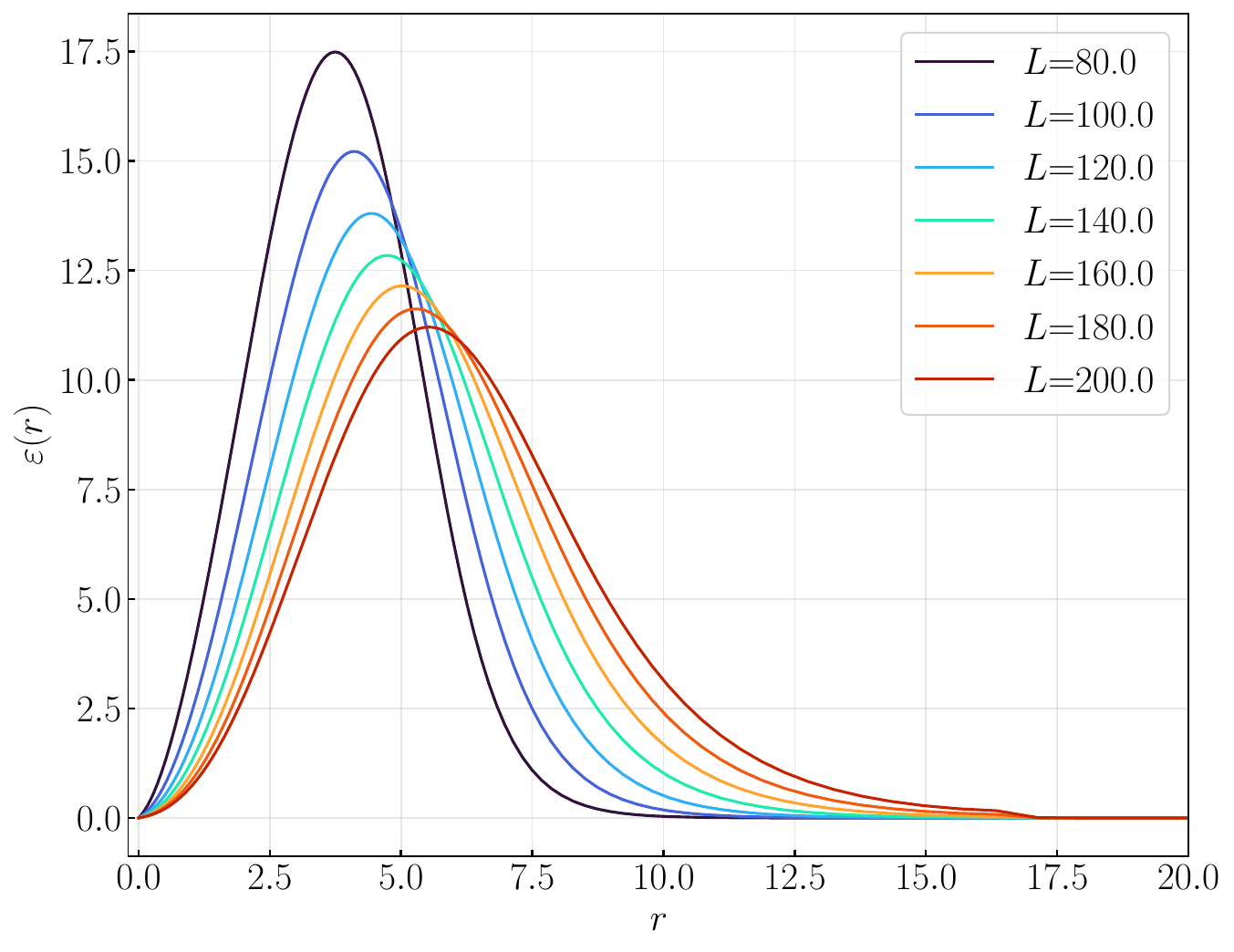}
    \vskip-2ex
    \caption{\footnotesize $\beta=20$}
    \label{fig:ED_beta20m10}
    \end{subfigure}
    \\[3mm]
    \begin{subfigure}[b]{0.47\textwidth}
    \centering
    \includegraphics[width=\linewidth]{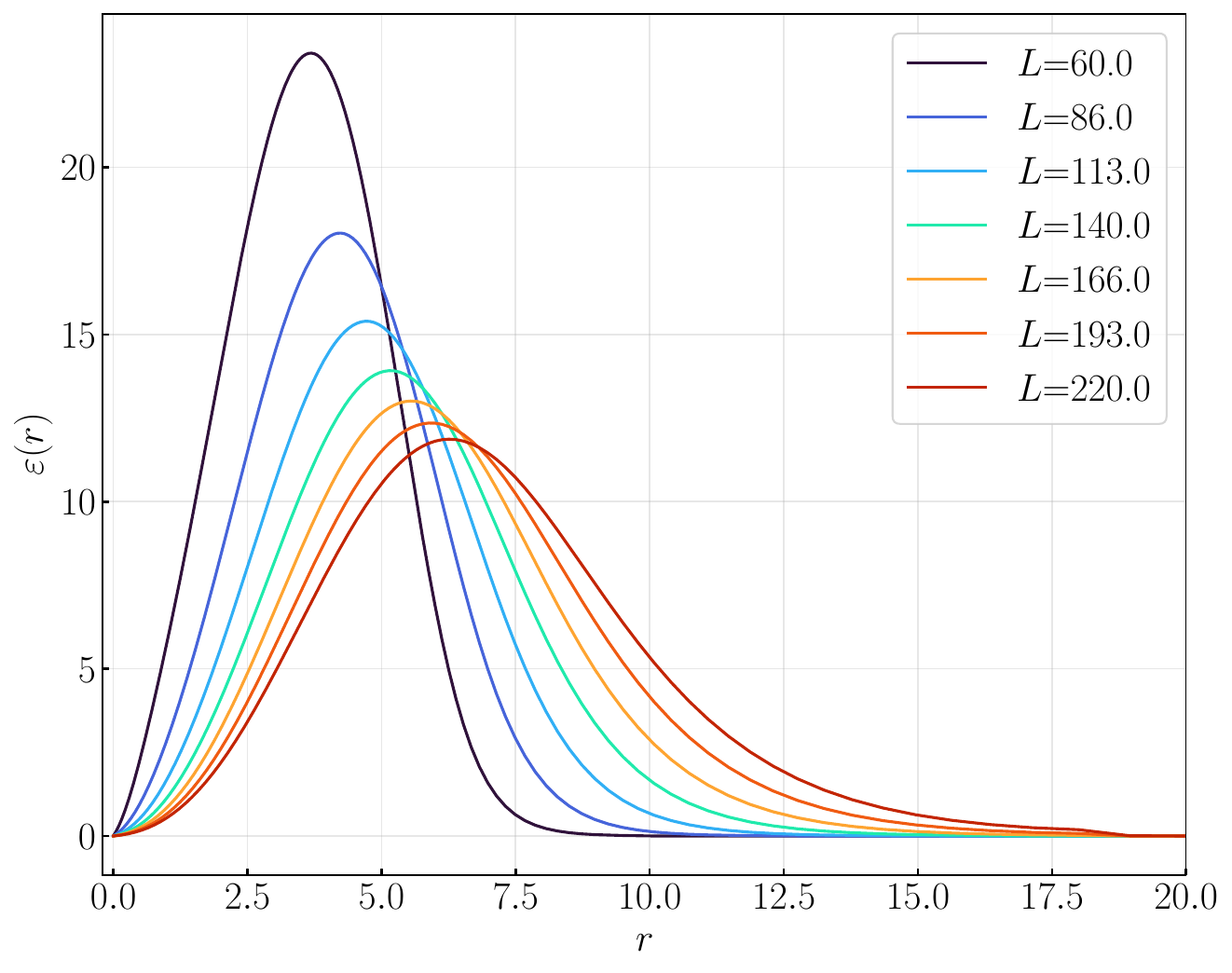}
    \vskip-2ex
    \caption{\footnotesize $\beta=25$}
    \label{fig:ED_beta25m10}
    \end{subfigure}
    \hfill
    \begin{subfigure}[b]{0.47\textwidth}
    \centering
    \includegraphics[width=\linewidth]{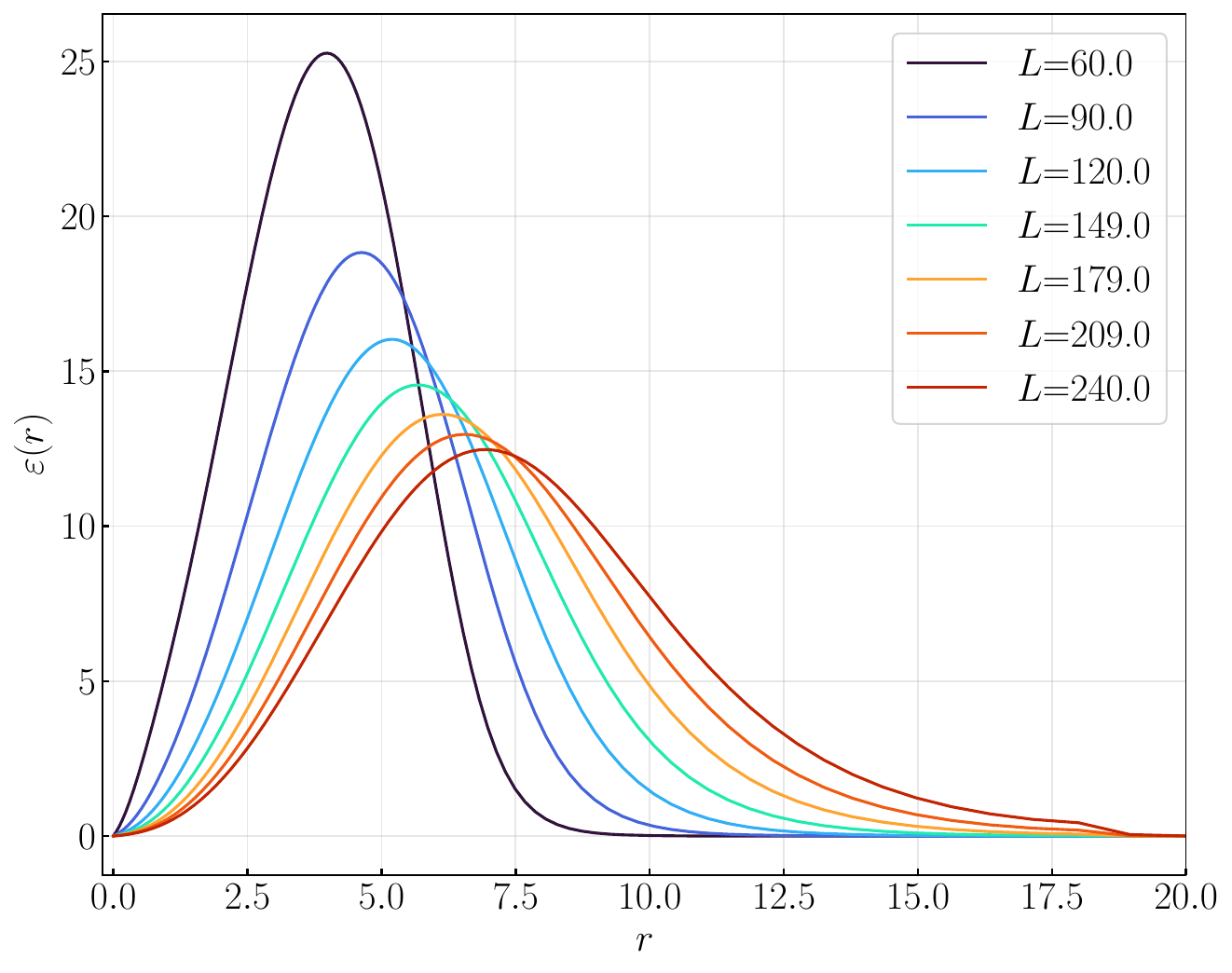}
    \vskip-2ex
    \caption{\footnotesize $\beta=30$}
    \label{fig:ED_beta30m10}
    \end{subfigure}
    \vskip-1ex
    \caption{\footnotesize String energy distribution over the radial coordinate for $n=1, g=1, m=10$}
    \label{fig:EDplots}
\end{figure}

The radial energy density $\varepsilon(r)$, defined as the integrand of \eqref{eq:energyr} integrated over  the angular variable and $z$  is shown in \cref{fig:EDplots}. The energy is concentrated around the string axis, with a peak at a finite radius $r_{\rm peak}$ that depends on both $\beta$ and $L$. This localized distribution illustrates that our solution is indeed a finite-energy soliton rather than a singular or delocalized configuration as is expected in the case of a twisted semilocal string. For all values of $\beta$, the energy density decays rapidly beyond the peak, falling to negligible values for $r \gtrsim 15$ to $20$.

A salient feature in \cref{fig:EDplots} is the systematic variation of the energy distribution with $L$. For smaller $L$, the peak energy density is higher and the distribution is more sharply localized, reflecting the concentration of the twist energy in the core. As $L$ increases toward $L_*$ and beyond, the peak height decreases and the distribution  slightly broadens. However, the rate of broadening becomes lower with increasing $L$, suggesting that the energy distribution saturates to a finite width. This ensures that our twisted  string remains a localized object (in the directions perpendicular to $z$) even for very large $L$.

The radial extent of the energy distribution also depends on $\beta$. Since the last term in \eqref{eq:energyr} is proportional to $(\beta - 1)$, larger values of $\beta$ lead to enhanced energy contribution due to sharp field gradients, causing the profile to spread over a wider radial region. 
As a result, the energy density peak shifts outward and the distribution becomes broader with increasing $\beta$.
This effect is most evident when comparing the $\beta = 10$ and $\beta = 30$ subplots in \cref{fig:EDplots}, where at the same string length $L$ the energy density peak is shifted to a larger radius for larger $\beta$.

Finally, we note that while so far we limited our analysis to the case $n = 1$ (i.e. unit winding in the azimuthal direction), the stability mechanism is expected to persist for higher winding numbers too.

The Hopf-like invariant $h = kn$, see \eqref{eq:hopflike}, would then take larger values, corresponding to more complicated knotted field configurations of higher-charge vortons.
The interplay between azimuthal winding $n$ and longitudinal twist $k$, as well as the resulting higher knotted structures, is left for a future investigation.

\section{Comparison with the literature \label{sec:compare}}

Based on our numerical results, we contrast our findings with previous numerical studies of Hopf and Hopf-like solitons and elucidate the connection to recent analytical works.

\subsection{Comparison with J\"aykk\"a and Speight}

The numerical study by J\"aykk\"a and Speight \cite{4} addressed the issue whether the knot solitons of the Faddeev-Skyrme model survive when the current coupling of the full two-component Ginzburg-Landau (TCGL) theory of a superconductor \cite{Babaev:2001zy} is restored. Their approach is to interpolate between the Faddeev-Skyrme model ($\eta=0$ in \cref{eq:JS_energy}) and the TCGL model in the
sigma-model limit ($\eta=1$ in \cref{eq:JS_energy}) via a one-parameter family of energy
functionals,
\begin{align}
E^{(\eta)} 
= \frac{1}{8}\norm{\dd{\phi}}^2 + \frac{1}{8}\norm{\phi^{*}\omega}^2
+ \frac{\eta}{2}\langle dC,\,\phi^*\omega\rangle
+ \frac{1}{2}\norm{\dd{C}}^2 + \frac{1}{2}\norm{C}^2\,,
\label{eq:JS_energy}
\end{align}
in which the coupling between the homogeneous coordinate $\phi$ and the current one-form $C$ is controlled by $\eta$. 
They track the energy minimizers from $\eta=0$ to $\eta=1$ and find that, for Hopf charges $\cH=1,\,2,\,3$, the soliton core shrinks monotonically and the energy approaches zero as $\eta\to 1$. 
The mechanism is identified with the Derrick-scaling instability. Namely, as $\eta$ increases, the minimizers approach the condition
\begin{equation}
dC + \tfrac{1}{2}\varphi^*\omega = 0\,,
\label{eq:JS_derrick_cond}
\end{equation}
under which no scale-fixing term survives and $E^{(1)} \to E^{(1)}/\lambda$, vanishing as $\lambda\to\infty$. Here $\lambda$ is the scaling constant, say, taking $x^{\mu}\to\lambda x^{\mu}$ .\\

Our construction differs from that of \cite{4} in several respects.

\paragraph{Parameter regime.}
While \cite{4} works strictly in the sigma-model limit $\beta\to\infty$, we study the energy functional \eqref{eq:energyr} at finite values of $\beta = 10,\,20,\,25,$ and $30$. 
At finite $\beta$, the fourth-order term proportional to $(\beta-1)$ in \cref{eq:energyr} provides a nontrivial contribution that competes with the quadratic terms, and the sigma-model truncation need not be a direct approximation.

\paragraph{Topological sector and geometry.}
The solitons studied in \cite{4} are point-like Hopfions in $\mathbb{R}^3$, where the Hopf invariant $\cH\in\pi_3(S^2)$ classifies maps from compactified three-space $S^3$ to $S^2$.
In contrast, our construction is based on a twisted semilocal string extended along the $z$ direction which is similar to considering the base space of $\mathbb{R}^2\times S^1$.
In our case, the relevant invariant is the Hopf-like charge $h=kn$ defined in \cref{eq:hopflike}.
The Hopf-like soliton here is obtained by bending this string into a torus, as depicted in \cref{fig1}, following the detailed analysis of \cite{3}. As noted at the end of the Introduction, when the loop length greatly exceeds the tube thickness the extrinsic-curvature-induced destabilizing force is parametrically small and the resulting toroidal configuration is at least quasi-stable.

There is a further distinction related to the Derrick scaling argument \cite{Hobart:1963elz,Derrick:1964ww}. 
Consider a simultaneous rescaling of all three spatial coordinates of a localized configuration. In our system, the energy \cref{eq:energyr} scales as
\begin{align}
  \cE(\lambda) = \lambda \cE_{(1)} + \frac{1}{\lambda}\cE_{(-1)}
\end{align}
where both $\cE_{(1)}$ and $\cE_{(-1)}$ are positive definite. The energy therefore possesses a minimum at a finite scaling factor $\lambda_*$, determined by $\dv*{\cE}{\lambda}|_{\lambda_*}=0$, and the Derrick-type destabilization does not occur. 
This is to be contrasted with the aforementioned situation in \cite{4}, where the energy diminishes in the limit $\lambda\to\infty$.

\paragraph{Role of the gauge field.}
In the formulation of \cite{4}, the supercurrent $C$ is treated as an independent dynamical field whose coupling to $\phi(r)$ is varied. 
The instability arises precisely because, at $\eta=1$, the minimizer adjusts $C$ to satisfy eq.~(\ref{eq:JS_derrick_cond}), thereby eliminating the stabilizing quartic terms (the middle three terms in \eqref{eq:JS_energy}). 
In our ansatz, the gauge field $A_i$ is fully determined by the scalar profile $\varphi(r)$ and the twist function $\alpha(z)$ through eqs.~(\ref{eq:Sinphi}) and~(\ref{eq:Ainphi}). 
To see the connection more explicitly, recall the ``supercurrent'' defined in \cite{4} (see also \cite{2}) by the formula
\begin{equation}
C = \frac{i}{2\xi}\sum_a\bigl(\phi_a\,d\phi_a
- \overline{\phi}_a\,d\phi_a\bigr) + A\,.
\label{eq:supercurrent_def}
\end{equation}
Within our semilocal-string ansatz, the gauge field
eq.~(\ref{eq:Ainphi}) is precisely the pure-gauge piece
$A = -\frac{i}{2\xi}\sum_a(\phi_a\,d\bar{\phi}_a
- \overline{\phi}_a\,d\phi_a)$,\footnote{Here $\phi_{a}$ is identical to the $n_{A}$ field in \cref{eq:nparam}. Same for their complex conjugates.} so that $C=0$ identically.
Consequently, the specific instability channel identified in \cite{4}, in which $C$ adjusts dynamically to satisfy \cref{eq:JS_derrick_cond} and eliminates the stabilizing quartic terms, is absent in our construction.

In light of the differences discussed above, our construction and that of \cite{4} operate in distinct regimes of two-flavor scalar QED.
We work with the low-energy effective theory at finite $\beta$, whereas \cite{4} focuses on the UV regime in the sigma-model limit. 
The two sets of conclusions are therefore not in contradiction but rather complementary to each other.

\subsection{Recent analytic development}

A closely related analytic construction was recently presented by Balakrishnan, Dandoloff, and Saxena \cite{5}, who studied Hopfion-vortex excitations in the continuum limit of an inhomogeneous XXZ Heisenberg ferromagnet. Their energy functional
\begin{align}\label{eq:EBDS}
E_{\rm BDS} &= \int\dd[3]{x} \left\{ 
    (\d_x\vec{S})^{2} + (\d_y\vec{S})^2 
    + \frac{\tilde{K}_3}{r^2} (\d_{z}\vec{S})^2 
\right\}
\nonumber\\[2mm]
&= \int\dd{r}\dd{\theta}\dd{z} \left\{  
  r(\d_r\Theta)^2 + \frac{n^2}{r}\sin^{2}{\Theta}
  + \frac{4\pi^2k^2\tilde{K}_3}{L^2r}\sin^{2}{\Theta}
\right\}
\,.
\end{align}
describes an O(3) sigma model in which the exchange coupling in the $z$-direction carries a spatial inhomogeneity. 
This judiciously chosen interaction makes the system exactly solvable with the ansatz 
\begin{align}\label{eq:Ssphtwist}
\vec{S} = \left( 
  \sin{\Theta(r)}\cos(n\theta+\frac{2\pi k z}{L}) \,,\,
  \sin{\Theta(r)}\sin(n\theta+\frac{2\pi k z}{L}) \,,\,
  \cos{\Theta(r)}
\right)
\end{align}
and, through a Derrick scaling argument \cite{Derrick:1964ww,Hobart:1963elz}, ensures the energetic stability of the resulting three-dimensional soliton against uniform rescaling of the coordinates.

The twisted skyrmion-string ansatz \eqref{eq:Ssphtwist} employed in \cite{5} is equivalent to the semilocal-string ansatz of \cref{eq:Sinphi,eq:Ainphi}.
In both cases the unit vector $\vec{S}$ winds $n$ times around the azimuthal angle and acquires a linear twist $\alpha(z) = 2\pi k z/L$ along the string axis. 
The topological content is likewise parallel. Reference \cite{5} considers the ordinary Hopf invariant while in our framework the Hopf-like charge defined in \cref{eq:hopflike}. 
Neither quantity is a Hopf invariant in the strict $\pi_3(S^2)$ sense because both constructions live on $\RR^2\times S^1$ rather than compactified $S^3$.
In addition this Hopfion string suggests the lower bound relation of the energy $E_{\rm BDS} \geq c_0 \cH^{1/2}$ (see for example \cite{manton,Shnir:2018yzp} for review) rather than a typical one $E \geq c_0 \cH^{3/4}$ of a Hopfion.

To make the connection between the two models precise, we recast our energy functional \eqref{eq:energy} in terms of the  spherical coordinate fields $(\Theta, \Phi)$\footnote{Note that $\Theta(r)$ can be related to $\vp(r)$ as
\begin{align}
    \sin{\Theta} = 2\vp\sqrt{1-\vp^2}
\end{align}
This is well-defined because it can be proven that $0 \leq 2\vp\sqrt{1-\vp^2} \leq \vp^2+(1-\vp^2) = 1$ and $\Theta\in[0,\pi)$ as anticipated.}
\begin{multline}\label{eq:energyS}
    \cE = \int_{0}^{L}\dd{z}\int_{\RR}\dd{r}\dd{\theta} \Biggl\{
    \left[ \frac{r}{4}(\d_r\Theta)^2
    + \frac{n^2}{4r}\sin^2{\Theta} \right]
    \\
    + \sin^{2}{\Theta} \cdot \left[  
        \frac{k^2\pi^2r}{L^2}
        + \frac{\beta-1}{8g^2}
        \left( \frac{n^2}{r} + \frac{4k^2\pi^2 r}{L^2} \right)(\d_r\Theta)^2
    \right]
    \Biggr\}
\end{multline}
The terms in the square bracket on the first line of Eq.~\cref{eq:energyS} are, up to an overall constant, identical to $(\partial_x\vec{S})^2 + (\partial_y\vec{S})^2$ of $E_{\rm BDS}$. 
The second line of \cref{eq:energyS} contains more nonlinear contributions proportional to $(\beta-1)/g^2$ that couple the twist $\alpha'(z)$ and the azimuthal winding $n$ to the radial gradient $(\partial_r\Theta)^2$.

Despite this structural difference, the two models becomes similar in a certin limit. Expanding the profile function near the string core (i.e. $r \to 0$) for $n=1$,\footnote{In general, the leading-order behavior of $\Theta(r)$ near the origin is $r^{|n|}$ to satisfy the corresponding equation of motion. We take $n=1$ here to match the winding number used in our numerical analysis.}
the leading order approximation of $\Theta(r)$ around the origin reads
\begin{align}
  \Theta(r) = a_1 r + \cdots
\end{align}
and terms with coefficients linear in $r$ can be ignored.
Then, $\cE$ can be approximated by $E_{\rm BDS}$ by the identification
\begin{align}
  \frac{4\pi^2k^2\tilde{K}_3}{L^2} \equiv \frac{\beta-1}{8g^2}a_1^2
  \,.
\end{align}
In this sense, the BDS model may be viewed as the small-$r$, leading-order approximation of our twisted semilocal vortex construction while our model constitutes a nonlinear generalization that incorporates the full quartic structure of the Faddeev--Skyrme theory.

\begin{figure}[t] 
  \centering 
  \includegraphics[width=.7\linewidth]{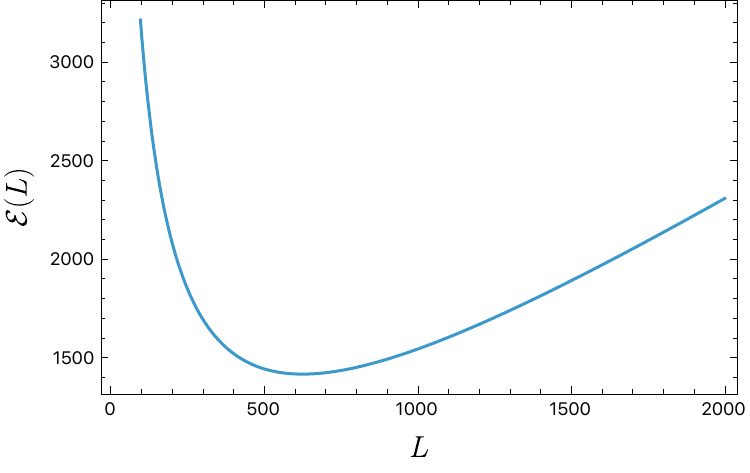}
  \vskip-3ex
  \caption{\footnotesize The estimation of the energy dependence on $L$ in \cite{3}. Here $\beta=30$, $k=10$, and $\abs{\rho}$ is taken to be $5$ as we found in \cref{fig:wvplots}.}
  \label{fig:GSYEL}
\end{figure}

The last point we would like to make in this section concerns the comparison of the $\cE(L)$ dependence between our numerical results and the analytic estimates in \cite{3} and \cite{5}.
Our numerical results verify the estimation of $\cE(L)$ given in eq. (5 .6) of \cite{3}, which is illustrated in \cref{fig:GSYEL}.
In the small-$L$ regime, the energy rises steeply, with the slope
\begin{align}
  \eval{\pdv{\cE}{L}}_{{\rm small}~L} \sim \frac{\xi\left( 2\pi^2\abs{\rho}k \right)}{L^2} \left( 1-\log\frac{L}{\abs{\rho}} \right)
\end{align}
diverging as $L \to 0$.
In the large-$L$ regime, the change of the energy with respect to the length approaches a constant,
\begin{align}
  \eval{\pdv{\cE}{L}}_{{\rm large}~L} \sim \xi + \frac{\beta}{3g^2\abs{\rho}^2}F\left( \frac{\beta}{\abs{\rho}^2m_{\gamma}^2} \right)
\end{align}
where $F(\kappa)$ is a fixed numerical factor for given model parameters.
The resulting $\cE(L)$ curve, with a divergent slope at small $L$ and a finite slope at large $L$, is consistent with the qualitativa picture shown in \cref{fig:ELplots}.

In the model of \cite{5}, Hopfion energy depends monotonically on $L$,
\begin{align}
  E(L) \sim \left( n^2L^2 + 4\pi^2k^2\tilde{K}_3 \right)^{1/2}
  \,,
\end{align}
exhibiting no minimum. 
However, this does not indicate an instability of the BDS construction.
In their setting, $L$ is a fixed external parameter, the thickness of a magnetic material sample, rather than a dynamical degree of freedom.
The stability of the construction given in \cite{5} follows from the scaling $(x,y,z) \to (\lambda x, \lambda y, \lambda z)$, which yields $E_{\rm BDS}(\lambda) = E_{1}\lambda + E_{-1}/\lambda$.
The competition between $E_{1}$ and $E_{-1}$ fixes the soliton at a definite scale $\lambda_1$.
The thickness $L$ then refers to the extent of the material at this scale.

\section{Discussions and conclusions}
\label{sec:cncld}
 
In this paper we have presented a numerical investigation of the classical stability of Hopf-like solitons, vortons,   realized as twisted semilocal vortex strings in the low-energy effective theory of two-flavor scalar QED.
Working on $\mathbb{R}^2 \times S^1$ with the energy functional \eqref{eq:energy} at finite values of the parameter~$\beta$, we found the numerical solution to the nonlinear profile equation \eqref{eq:eomvarphi} and traced the total energy $\mathcal{E}(L)$ as a function of the string length $L$.
For all values of $\beta$ examined ($\beta = 10,\, 20,\, 25,\, 30$), the energy exhibits a well-defined minimum at a finite equilibrium length $L_*$, indicating that the twisted vortex tube is dynamically stabilized as predicted in \cite{3}.
The equilibrium length $L_*$ and the soliton mass $M_{\mathrm{sol}} = \mathcal{E}(L_*)$ both grow monotonically with $\beta$, the scale hierarchy $L_* \gg |\rho| \gg 1/m_\gamma$ is confirmed by the numerical profiles as emphasized for the validity of the analytic estimates in \cite{3}.
The vorton charge $h = 10$ is preserved to $O(10^{-4})$ accuracy across all solutions.

Because the vorton charge $h = kn$ is a genuine topological quantum number, the vorton is not only a local energy minimum but is also protected against topological decay within the $\RR^2 \times S^1$ setup. 
When such a vorton is bent into a closed ring of circumference $L_\ell \gg L_s$, the extrinsic curvature is $O(1/L_\ell)$ and any curvature-induced destabilizing force is correspondingly small. Thus, even if this residual force eventually drives the toroidal configuration off strict classical stability on $\RR^3$, this can happen only through quantum tunneling. The resulting Hopf-like configuration remains quasi-stable and long-lived.
This provides a concrete physical ground for the conjecture of Faddeev and Noemi in the geometry considered here.

Our results are compared in detail with the recent numerical study \cite{4}, showing that Hopf solitons are destabilized by the ``supercurrent'' coupling in the two-component Ginzburg--Landau theory.
As discussed in Sec. \ref{sec:compare}, the two analyses  correspond to distinct regimes. Namely, \cite{4} works in the sigma-model limit ($\beta \to \infty$) on $\RR^3$ with a dynamical current $C$, while in our construction the gauge field is related to the scalar profile through the ansatz \eqref{eq:Sinphi}, forcing $C = 0$ identically.
The instability channel identified in \cite{4} is therefore absent, and our Derrick scaling analysis guarantees a minimum at finite scale $L_{*}$.
On the analytic side, our numerical $\mathcal{E}(L)$ curves are fully consistent with the estimation given in \cite{3}.

Several directions remain open.
The extension to higher azimuthal winding numbers $n > 1$ would probe configurations with larger Hopf-like charge and more intricate knotted structures.\footnote{
We have also carried out a preliminary test for the case $n=2$, which exhibits qualitatively similar stabilization behavior. However, the numerical solution is more susceptible to breakdown at higher winding number, and hence a further refinement of the solver is required before a systematic higher-charge analysis can be undertaken.}
A full three-dimensional simulation on $\RR^3$ would test whether the stabilization persists when a bent twisted string is embedded in $\RR^3$ \cite{Ireson:2018bdw}, and would make quantitative the quasi-stability argument based on small extrinsic curvature sketched above.
Lastly, this study can be extended to the deformed Faddeev--Skyrme model. As shown in~\cite{FOZ,Sheu:2023hoz,Gamayun:2023atu}, the target space of the deformed model is an elongated sphere that is topologically equivalent to the ordinary $S^2$.
Therefore, one expects the same $\pi_3(S^2) \cong \ZZ$ classification to persist and Hopfion or vorton excitations to exist within the resulting one-parameter family of models. However, the stabilization mechanism may be considerably richer by virtue of the additional nonlinear interactions governed by the deformation parameter $\kappa$.

\section*{Acknowledgements}

The work of MS is supported in
part by U.S. Department of Energy Grant No. de-sc0011842. CHS was supported in part by the US National Science Foundation under grant PHY-2210283.


\newpage

\begin{thebibliography}{99}

\bibitem{FN}
First mention of Hopfion see in Sec. 4 of L.D. Faddeev, {\em Quantization of solitons}, Preprint IAS Print-75-QS70 (Inst. Advanced Study, Princeton, NJ, 1975), 32 pp.;
L.~D.~Faddeev and A.~J.~Niemi,
{\em Knots and particles,}
Nature \textbf{387}, 58 (1997)
[arXiv:hep-th/9610193 [hep-th]]; {\em Toroidal configurations as stable solitons},
[arXiv:hep-th/9705176 [hep-th]].

\bibitem{3}
A.~Gorsky, M.~Shifman and A.~Yung,
{\em Revisiting the Faddeev-Skyrme Model and Hopf Solitons,}
Phys. Rev. D \textbf{88}, 045026 (2013)
[arXiv:1306.2364 [hep-th]].

\bibitem{2}
E.~Babaev,
{\em Non-Meissner electrodynamics and knotted solitons in two-component superconductors,}
Phys. Rev. B \textbf{79}, 104506 (2009)
[arXiv:0809.4468 [cond-mat.supr-con]].

\bibitem{Davis:1988jp}
R.~L.~Davis and E.~P.~S.~Shellard,
{\em The Physics of Vortex Superconductivity,}
Phys. Lett. B \textbf{207}, 404-410 (1988)

\bibitem{Davis:1988jq}
R.~L.~Davis and E.~P.~S.~Shellard,
{\em The Physics of Vortex Superconductivity. 2,}
Phys. Lett. B \textbf{209}, 485-490 (1988)

\bibitem{Radu:2008pp}
E.~Radu and M.~S.~Volkov,
{\em Existence of stationary, non-radiating ring solitons in field theory: knots and vortons,}
Phys. Rept. \textbf{468}, 101-151 (2008)
[arXiv:0804.1357 [hep-th]].



\bibitem{manton} 
N. Manton and P. Sutcliffe,  {\em Topological
Solitons}, (Cambridge University Press, 2004), Sect. 9.11.

 \bibitem{nsvzrev}
 V.~A.~Novikov, M.~A.~Shifman, A.~I.~Vainshtein and V.~I.~Zakharov,
{\em Two-Dimensional Sigma Models: Modeling Nonperturbative Effects Of Quantum
  Chromodynamics},
  Phys.\ Rept.\  {\bf 116}, 103 (1984).
  
  \bibitem{uch} 
   M. Shifman, {\em Advanced Topics in Quantum Field Theory}, 2$^{nd}$ Edition,
   (Cambridge University Press, 2020).
  
  

  
\bibitem{Shnir:2018yzp}
Y.~M.~Shnir,
{\em Topological and Non-Topological Solitons in Scalar Field Theories,}
(Cambridge University Press, 2018).


\bibitem{4}
J.~J\"aykk\"a and J.~M.~Speight,
{\em Supercurrent coupling destabilizes knot solitons,}
Phys. Rev. D \textbf{84}, 125035 (2011)
[arXiv:1106.5679 [math-ph]].





\bibitem{5}
R.~Balakrishnan, R.~Dandoloff and A.~Saxena,
{\em Exact hopfion vortices in a 3D Heisenberg ferromagnet,}
Phys. Lett. A \textbf{480}, 128975 (2023)
[arXiv:2202.07195 [nlin.SI]].


\bibitem{Ireson:2018bdw}
E.~Ireson, M.~Shifman and A.~Yung,
{\em Supersymmetrizing the Gorsky-Shifman-Yung soliton,}
Phys. Rev. D \textbf{97}, no.10, 105021 (2018)
[arXiv:1803.04549 [hep-th]].

\bibitem{Babaev:2001zy}
E.~Babaev, L.~D.~Faddeev and A.~J.~Niemi,
{\em Hidden symmetry and knot solitons in a charged two-condensate Bose system,}
Phys. Rev. B \textbf{65}, 100512 (2002)
[arXiv:cond-mat/0106152 [cond-mat.supr-con]].

\bibitem{Hobart:1963elz}
R.~H.~Hobart,
{\em On the Instability of a Class of Unitary Field Models,}
Proc. Phys. Soc. \textbf{82}, no.2, 201 (1963)


\bibitem{Derrick:1964ww}
G.~H.~Derrick,
{\em Comments on nonlinear wave equations as models for elementary particles,}
J. Math. Phys. \textbf{5}, 1252-1254 (1964)

\bibitem{FOZ}
V.~A.~Fateev, E.~Onofri and A.~B.~Zamolodchikov,
{\em Integrable deformations of the $O(3)$ sigma model. The sausage model},
Nucl. Phys. B \textbf{406}, 521-565 (1993).

\bibitem{Sheu:2023hoz}
C.~H.~Sheu and M.~Shifman,
{\em Remarks on baby Skyrmion Lie-algebraic generalization,}
Phys. Rev. D \textbf{108}, no.6, 065003 (2023)
[erratum: Phys. Rev. D \textbf{110}, no.2, 029901 (2024)]
[arXiv:2303.12597 [hep-th]].

\bibitem{Gamayun:2023atu}
O.~Gamayun, A.~Losev and M.~Shifman,
{\em Peculiarities of beta functions in sigma models,}
JHEP \textbf{10}, 097 (2023)
[arXiv:2307.04665 [hep-th]].



\end{thebibliography}
\end{document}